\documentclass[%
preprint,
showpacs,
 amsmath,amssymb,
 aps,
floatfix,
]{revtex4-1}

\usepackage{graphicx}
\newcommand{\bi}{\mathbf}
\usepackage{bm}
\usepackage{hyperref}

\renewcommand{\d}{\,\mathrm{d}}

\def\bra#1{\langle#1\vert}
\def\ket#1{\vert#1\rangle}
\def\braket#1#2{\langle #1 \vert #2 \rangle}

\usepackage{color}

\begin{document}

\preprint{}

\title{Light-matter interaction in the long-wavelength limit: no ground-state without dipole self-energy}
\author{ Vasil Rokaj}
\email{vasil.rokaj@mpsd.mpg.de}
\author{Davis M. Welakuh}
\email{davis.welakuh@mpsd.mpg.de}
\author{Michael Ruggenthaler} 
\email{michael.ruggenthaler@mpsd.mpg.de}
\author{Angel Rubio}
\email{angel.rubio@mpsd.mpg.de}
\affiliation{\footnotesize Department of Physics, Max Planck Institute for the Structure and Dynamics of Matter and Center for Free-Electron Laser Science, 22761 Hamburg, Germany}

\date{\today}

\begin{abstract}
Most theoretical studies for correlated light-matter systems are performed within the long-wavelength limit, i.e., the electromagnetic field is assumed to be spatially uniform. In this limit the so-called length-gauge transformation for a fully quantized light-matter system gives rise to a dipole self-energy term in the Hamiltonian, i.e., a harmonic potential of the total dipole matter moment. In practice this term is often discarded as it is assumed to be subsumed in the kinetic energy term. In this work we show the necessity of the dipole self-energy term. First and foremost, without it the light-matter system in the long-wavelength limit does not have a ground-state, i.e., the combined light-matter system is unstable. Further, the mixing of matter and photon degrees of freedom due to the length-gauge transformation, which also changes the representation of the translation operator for matter, gives rise to the Maxwell equations in matter and the omittance of the dipole self-energy leads to a violation of these equations. Specifically we show that without the dipole self-energy the so-called ``depolarization shift'' is not properly described. Finally we show that this term also arises if we perform the semi-classical limit \textit{after} the length-gauge transformation. In contrast to the standard approach where the semi-classical limit is performed \textit{before} the length-gauge transformation, the resulting Hamiltonian is bounded from below and thus supports ground states. This is very important for practical calculations and for density-functional variational implementations of the non-relativistic QED formalism. For example, the existence of a combined light-matter ground state allows to calculate the Stark shift non-perturbatively.
\end{abstract}

\maketitle


\section{\label{Sec:1} Introduction}

Todays fundamental description of charged particles interacting with electromagnetic fields is based on quantum-electrodynamics (QED)~\cite{Cohen,ryder1996, greiner2013}. In the context of high-energy physics, where the kinetic energy of the particles dominates, QED is usually formulated in terms of a perturbative scattering theory and has been proven to be exceptionally accurate~\cite{ryder1996, greiner2013}. If we are interested in the properties of molecular or solid-state systems, we are mainly dealing with the low-energy limit of QED~\cite{craig1984, spohn2004, Ruggi PRA 2014} and an approximate description of the matter system in terms of the many-body Schr\"odinger equation is usually sufficient. In this case the effect of the photons on the matter degrees of freedom is encoded in the Coulomb interaction, which neglects the transversal photon degrees of freedom~\cite{craig1984, spohn2004, Ruggi PRA 2014}. On the other hand, in the context of quantum optics and photonics~\cite{grynberg2010}, the transversal degrees of the electromagnetic field are the central object of interest and rather the matter degrees of freedom are approximated. A common strategy is to only keep a few states that describe the matter system and then to restrict the participating modes of the photon field, which leads to a few-level-few-modes approximation, which in its simplest form is known as the Rabi or Jaynes-Cummings model~\cite{grynberg2010, Shore JMO 1993, Braak}. A different approach is to use the macroscopic Maxwell equations~\cite{jackson1975} and to employ linear-response functions to fix the constitutive relations, i.e., the dependence of the polarization and magnetization on the external fields.

Recent experimental advances~\cite{you-2011, schwartz-2011, Hutchison2012, shalabney-2015a, shalabney-2015b, chikkaraddy-2016, ebbesen-2016} at the interface between quantum chemistry, solid-state physics and quantum optics have uncovered situations where the above usual theoretical simplifications of the full QED description become questionable~\cite{Johannes PNAS 2017, ruggenthaler2017, George}. In such situations matter is strongly-coupled to photons which gives rise to novel phenomena such as changes in chemical reaction rates~\cite{Galego2016, galego2017}, appearance of attractive photons due to correlated matter-photon states~\cite{firstenberg2013} or an increase in conductivity due to photon-matter hybridization~\cite{orgiu-2015}.  For such experiments to be described theoretically, light and matter have to be treated on equal, quantized footing. Such a balanced description that reduces in the different limits to the many-body Schr\"{o}dinger equation on the one hand and to the usual model Hamiltonians of quantum optics on the other is non-relativistic QED~\cite{spohn2004}. So far non-relativistic QED has not been applied much in practice due to its inherent computational complexity, but recent approaches such as the exact formal density-functional reformulation of QED~\cite{ruggenthaler-2011, tokatly-2013, Ruggi PRA 2014}, generalized Green's function methods~\cite{melo2015} or extensions of quantum-chemical approximations~\cite{Johannes JCTC 2017}, make non-relativistic QED computationally feasible and already first calculations have been performed~\cite{Johannes PNAS 2015, Johannes PNAS 2017}.
 
To describe the above experimental situations, non-relativistic QED can be simplified further. In situations such as those of cavity and circuit QED~\cite{walther-2006,schmidt2013} the long-wavelength limit (dipole approximation) is implemented since in most cases the wave-length of the relevant electromagnetic modes is much larger than the spatial extension of the molecular system. The long-wavelength limit is the starting point in most investigations in quantum optics and molecular physics. It has two equivalent formulations, which are called the velocity and the length gauge, respectively~\cite{Cohen, Faisal}. In the velocity gauge, which is just the minimal-coupling prescription for a uniform vector potential, the square of the vector-potential operator appears. This term can be absorbed by a change in the frequency and the polarization direction of the electromagnetic field~\cite{Faisal}. A different way to avoid the square of the vector-potential operator is to make a unitary transformation and work in the length-gauge picture (see equation~(\ref{eq2.9})). This transformation leads to the appearance of the so-called dipole self-energy term (see equation~(\ref{eq2.15})), which is proportional to the square of the total-dipole operator. The dipole self-energy, however, cannot be absorbed by a change of frequency and polarization. Nevertheless, this term is often considered unimportant and is neglected with arguments based on the thermodynamic limit in the mean-field picture~\cite{Faisal}. Moreover, in the case of the Rabi and Jaynes-Cummings model this dipole self-energy term is just a constant energy offset and thus usually discarded and often taken synonymously with the square of the vector potential of the electromagnetic field~\cite{Vukics PRL, Vukics}. However, recent experimental and numerical results~\cite{Todorov PRL 2010, Todorov PRB 2012, Todorov PRX 2014, Vasanelli CRP 2016} highlight the importance of the dipole self-energy for non-relativistic QED in the long-wavelength limit. For instance, in Ref.~\cite{George} the polaritonic dispersion inside a cavity was measured and it was found that the dipole self-energy had to be taken into account to fit with the experimental data and in Ref.~\cite{Johannes PNAS 2017} some of us have shown numerically that without the dipole self-energy the electron-nuclear-photon system becomes unstable for ultra strong-coupling situations.

In this work we show mathematically the necessity of the dipole self-energy term and its physical significance for coupled matter-photon systems. More specifically, we show that no ground-state exists without this term. We highlight how this term affects the translational invariance and that it is necessary for the electric field to comply with the Maxwell equations in matter. The dipole self-energy gives rise to the plasma frequency that subsequently leads to the so-called depolarization shift. We further show the appearance of the dipole self-energy in the semi-classical limit which guarantees that a system subject to a static external electric field is bounded from below. This makes it possible, e.g., to treat the Stark shift non-perturbatively.

\section{\label{sec:2} Non-relativistic quantum-electrodynamics in the long-wavelength limit}

In the following we will consider the low-energy limit of QED, which is supposed to be sufficient to capture most effects encountered in atomic, molecular and solid-state systems~\cite{Cohen, craig1984, spohn2004}. For simplicity we will employ the clamped-nuclei approximation, i.e., that we treat the nuclei as fixed external attractive potentials, and consider electron-photon systems. The extension to electron-nuclei-photon systems is straightforward and all results apply equally to this more general case. Our starting point is the so-called Pauli-Fierz Hamiltonian~\cite{spohn2004} that describes electrons minimally coupled with photons~\cite{Cohen, Ruggi PRA 2014, Faisal}
\begin{eqnarray}\label{eq2.1} 
\hat{H}&=&\frac{1}{2m}\sum\limits^{N}_{j=1}\left(i\hbar \mathbf{\nabla}_{j}+\frac{e}{c} \hat{\mathbf{A}}(\bi{r}_j)\right)^2 +\frac{1}{4\pi\epsilon_0}\sum\limits^{N}_{j< k}\frac{e^2}{|\mathbf{r}_j-\mathbf{r}_k|}+\sum\limits^{N}_{j=1}v_{ext}(\mathbf{r}_{j})\nonumber\\
&+&\sum\limits_{\bi{n},\lambda}\left[-\frac{\hbar \omega_{n}}{2}\frac{\partial^2}{\partial q^2_{\bi{n},\lambda}} +\frac{\hbar\omega_{n}}{2}q^2_{\bi{n},\lambda}\right] .
\end{eqnarray}
where we neglected the Pauli (Stern-Gerlach) term, i.e., $\hat{\bm{\sigma}}\cdot\hat{ \bi{B}}(\bi{r})$, where $\hat{\bm{\sigma}}$ is a vector of the standard Pauli matrices and $\hat{\bi{B}}(\bi{r})$ corresponds to the magnetic field, since it will not contribute in the long-wavelength limit. Here $m$ is the mass of the electron, $\hbar$ the reduced Planck constant, $e$ is the charge of the electron, $c$ the velocity of light and $\hat{\mathbf{A}}(\bi{r})$ is the quantized vector potential of the electromagnetic field in Coulomb gauge~\cite{greiner2013}, given by the expression
\begin{equation}\label{eq2.4b}
\hat{\bi{A}}(\bi{r})=\left(\frac{\hbar c^2}{\epsilon_0 L^3}\right)^{\frac{1}{2}}\sum_{\bi{n},\lambda}\frac{\bm{\epsilon}_{\bi{n},\lambda}}{\sqrt{2\omega_n}}\left[ \hat{a}_{\bi{n},\lambda}e^{i\bi{k}_n\cdot\bi{r}}+\hat{a}^{\dagger}_{\bi{n},\lambda}e^{-i\bi{k}_n\cdot\bi{r}}\right].
\end{equation}
Further, $\omega_{n}=c|\mathbf{n}|(2\pi/L)$ are the allowed frequencies in a quantization box of length $L$, $\epsilon_0$ the vacuum permittivity, $\lambda$ the two transversal polarization directions and $\bm{\epsilon}_{\bi{n},\lambda}$ are the transversal polarization vectors of each photon mode which are perpendicular to the direction of propagation $\bi{k}_n$~\cite{greiner2013}. The operators $\hat{a}_{\bi{n},\lambda}$ and $\hat{a}^{\dagger}_{\bi{n},\lambda}$ are annihilation and creation operators, respectively, and obey the usual commutation relations
\begin{equation}\label{eq2.4a}
	[\hat{a}_{\bi{n},\lambda},\hat{a}^{\dagger}_{\bi{m},\kappa}]=\delta_{\bi{n}\bi{m}}\delta_{\lambda\kappa}.
\end{equation}
The second term of equation~(\ref{eq2.1}) is the electron-electron interaction due to the longitudinal part of the Coulomb-gauged photon field for $L \rightarrow \infty$. The third term corresponds to an external scalar potential that acts on the electrons, e.g., the attractive potential of the nuclei. The last term of the Hamiltonian~(\ref{eq2.1}) corresponds to the energy of the photon field. This can be deduced from the definition of the electromagnetic-energy density which is the electric field squared plus the magnetic field squared. With the corresponding operators
\begin{eqnarray}
	\hat{\bi{E}}(\bi{r}) &=& \left(\frac{\hbar }{\epsilon_0 L^3}\right)^{\frac{1}{2}}\sum_{\bi{n},\lambda} i\sqrt{\frac{\omega_{n}}{2}}\bm{\epsilon}_{\bi{n},\lambda} \left[\hat{a}_{\bi{n},\lambda}e^{i\bi{k}_{n}\cdot\bi{r}} - \hat{a}_{\bi{n},\lambda}^{\dagger}e^{-i\bi{k}_{n}\cdot\bi{r}} \right], \label{eq2.17}\\
	\hat{\bi{B}}(\bi{r}) &=& \left(\frac{\hbar }{\epsilon_0 L^3}\right)^{\frac{1}{2}}\sum_{\bi{n},\lambda} \frac{i\bi{k}_n\times\bm{\epsilon}_{\bi{n},\lambda}}{\sqrt{2 \omega_{n}}}\left[\hat{a}_{\bi{n},\lambda}e^{i\bi{k}_{n}\cdot\bi{r}} - \hat{a}_{\bi{n},\lambda}^{\dagger}e^{-i\bi{k}_{n}\cdot\bi{r}} \right],\label{eq2.18}
\end{eqnarray}
the energy becomes
\begin{equation}\label{eq2.3a}
    \hat{H}_p=\frac{\epsilon_0}{2}\int\limits_{L^3}\mathrm{d}^3 r\left[\hat{\bi{E}}^2(\bi{r})+c^2\hat{\bi{B}}^2(\bi{r})\right]=\sum_{\bi{n},\lambda}\hbar\omega_{n}\left[\hat{a}^{\dagger}_{\bi{n},\lambda}\hat{a}_{\bi{n},\lambda}+\frac{1}{2}\right].
\end{equation}
By introducing the displacement coordinates $q_{\bi{n},\lambda}$ and their conjugate momenta $\partial/\partial q_{\bi{n},\lambda}$ we can define the annihilation and creation operators as
 \begin{equation}\label{eq2.3}
	\hat{a}_{\bi{n},\lambda}=\frac{1}{\sqrt{2}}\left(q_{\bi{n},\lambda}+\frac{\partial}{\partial q_{\bi{n},\lambda}}\right)\qquad \textrm{and} \qquad \hat{a}^{\dagger}_{\bi{n},\lambda}=\frac{1}{\sqrt{2}}\left(q_{\bi{n},\lambda}-\frac{\partial}{\partial q_{\bi{n},\lambda}}\right).
\end{equation}
Substituting these expressions into equation (\ref{eq2.3a}) we obtain the equivalent form of the photon Hamiltonian
\begin{equation}\label{eq2.2}
	\hat{H}_p=\sum_{\bi{n},\lambda}\hbar\omega_{n}\left[\hat{a}^{\dagger}_{\bi{n},\lambda}\hat{a}_{\bi{n},\lambda}+\frac{1}{2}\right]=\sum_{\bi{n},\lambda}\left[-\frac{\hbar \omega_{n}}{2}\frac{\partial^2}{\partial q^2_{\bi{n},\lambda}} +\frac{\hbar\omega_{n}}{2}q^2_{\bi{n},\lambda}\right].
\end{equation}
Assuming a form factor for the photon modes, i.e., a square integrable mask function that suppresses infinitely high photon frequencies, and allowing external fields of Kato type the Pauli-Fierz Hamiltonian is bounded from below and thus obeys a variational principle for ground states~\cite{spohn2004} (see Sec.~\ref{sec:3} for details). We note that if we replaced the non-relativistic kinetic energy operator by the fully relativistic Dirac momentum, the resulting Hamiltonian would no longer be bounded from below and thus have no ground state. This is the main reason to work in the non-relativistic limit of QED. However, in practice a simplified form of the Pauli-Fierz Hamiltonian is used when describing, e.g., atoms or molecules in an optical cavity or subject to an optical laser pulse (described still fully quantized). In this case the wavelength of the relevant photon modes is much larger than the size of the electronic system, and we can neglect the spatial variation of the electromagnetic field $e^{\pm i\bi{k}_n\cdot\bi{r}}\approx 1$ (see Sec.~\ref{sec:3} for a more precise statement). This approximation is known by different names: it is either called the long-wavelength or optical limit as well as dipole approximation~\cite{Cohen, Faisal, spohn2004}. If we restrict ourselves to arbitrarily many but a finite number $M$ of modes $\alpha\equiv (\bi{n},\lambda)$\;, the vector potential $\hat{\mathbf{A}}(\bi{r})$ in this limit is given by
\begin{equation}\label{eq2.5}
	\hat{\mathbf{A}}=\sum^{M}_{\alpha=1}\frac{C \bm{\epsilon}_{\alpha}}{\sqrt{\omega_{\alpha}}}q_{\alpha} \qquad \textrm{where} \qquad C=\left(\frac{\hbar c^2}{\epsilon_0 L^3}\right)^{\frac{1}{2}}.
\end{equation}
As a consequence the full matter-photon Hamiltonian of equation (\ref{eq2.1}) in the long-wavelength approximation takes the following form
\begin{eqnarray}\label{eq2.6}
	\hat{H}_V&=&\frac{1}{2m}\sum\limits^{N}_{j=1}\left[-\hbar^2\mathbf{\nabla}^2_{j} +2i\frac{e\hbar}{c} \hat{\mathbf{A}}\cdot\mathbf{\nabla}_j +\frac{e^2}{c^2} \hat{\mathbf{A}}^2\right]+\frac{1}{4\pi\epsilon_0}\sum\limits^{N}_{j< k}\frac{e^2}{|\mathbf{r}_j-\mathbf{r}_k|}\nonumber\\
	&+&\sum\limits^{N}_{j=1}v_{ext}(\mathbf{r}_{j})+\sum\limits^{M}_{\alpha=1}\left[-\frac{\hbar \omega_{\alpha}}{2}\frac{\partial^2}{\partial q^2_{\alpha}} +\frac{\hbar\omega_{\alpha}}{2}q^2_{\alpha}\right],
\end{eqnarray}
In this dipole approximation the square of the vector potential can be eliminated by introducing new frequencies and polarizations for the photon modes~\cite{Faisal}. We will not implement these changes to eliminate $\hat{\bi{A}}^2$ since this term will disappear in the following. We proceed by performing a unitary transformation, which is called the length-gauge transformation~\cite{Cohen, Faisal, spohn2004} and it is defined as 
\begin{equation}\label{eq2.7}
\hat{H}_{L}' = \hat{U}^{\dagger} \hat{H}_{V} \hat{U},  \qquad \hat{U}=\exp[\frac{i}{\hbar}\frac{e}{c}\hat{\mathbf{A}}\cdot \mathbf{R}],
\end{equation}
where $\bi{R}=\sum\limits^{N}_{i=1}\bi{r}_i$ is the total dipole operator. The individual terms in the Hamiltonian~(\ref{eq2.6}) transform as
\begin{eqnarray}\label{eq2.8}
&&2i\frac{\hbar e}{c}\hat{\mathbf{A}}\cdot \mathbf{\nabla}_i  \longrightarrow  2i\frac{\hbar e}{c}\hat{U}^{\dagger}\hat{\mathbf{A}}\cdot \mathbf{\nabla}_i \hat{U}= 2i\frac{\hbar e}{c}\hat{\mathbf{A}}\cdot \mathbf{\nabla}_i -\frac{2e^2}{c^2}\hat{\mathbf{A}}^{2},\nonumber\\
	&&-\hbar^2\mathbf{\nabla}^2_i \longrightarrow  -\hbar^2 \hat{U}^{\dagger}\mathbf{\nabla}^2_i  \hat{U}=-\hbar^2 \mathbf{\nabla}^2_i -2i\frac{\hbar e}{c}\hat{\mathbf{A}}\cdot \mathbf{\nabla}_i +\frac{e^2}{c^2}\hat{\mathbf{A}}^{2},\nonumber\\
	&&\frac{e^2}{c^2}\hat{\mathbf{A}}^{2} \longrightarrow \frac{e^2}{c^2}\hat{U}^{\dagger}\hat{\mathbf{A}}^2\hat{U}=\frac{e^2}{c^2}\hat{\mathbf{A}}^{2},\\
	&&-\frac{\partial^2}{\partial q^2_{\alpha}} \longrightarrow  -\hat{U}^{\dagger}\frac{\partial^2}{\partial q^2_{\alpha}}\; \hat{U}=-\frac{\partial^2}{\partial q^2_{\alpha}}-i\frac{2eC\bm{\epsilon}_{\alpha}\cdot\mathbf{R}}{\hbar c\sqrt{\omega_{\alpha}}}\frac{\partial}{\partial q_{\alpha}} +\left(\frac{eC\bm{\epsilon}_{\alpha}\cdot \mathbf{R}}{\hbar c\sqrt{\omega_{\alpha}}}\right)^2 \nonumber.
\end{eqnarray}
The other terms of the Hamiltonian (\ref{eq2.6}) are invariant since they commute with the operator $\hat{U}$\;. The Hamiltonian after the length gauge transformation looks as follows:
\begin{eqnarray}\label{eq2.9}
	\hat{H}^{'}_L&=&-\frac{\hbar^2}{2m}\sum\limits^{N}_{i=1}\bi{\nabla}^2_{i}+\frac{1}{4\pi\epsilon_0}\sum\limits^{N}_{i< j}\frac{e^2}{|\bi{r}_i-\bi{r}_j|}+\sum\limits^{N}_{i=1}v_{ext}(\bi{r}_{i})\nonumber\\
	&+&\sum\limits^{M}_{\alpha=1}\left[-\frac{\hbar\omega_{\alpha}}{2}\frac{\partial^2}{\partial q^2_{\alpha}}+\frac{\hbar\omega_{\alpha}}{2} q^2_{\alpha} -i\frac{\sqrt{\omega_{\alpha}}eC\bm{\epsilon}_{\alpha}\cdot\bi{R}}{ c}\frac{\partial}{\partial q_{\alpha}} +\frac{\hbar\omega_{\alpha}}{2}\left(\frac{eC\bm{\epsilon}_{\alpha}\cdot\bi{R}}{\hbar c\sqrt{\omega}_{\alpha}}\right)^2\right].
\end{eqnarray}
We thus see that the square of the vector potential has been exactly eliminated after the length gauge transformation and a new term appears in the Hamiltonian, an electronic harmonic potential $\left(\bm{\epsilon}_{\alpha}\cdot\bi{R}\right)^2$\;. This is the dipole self-energy term. But, as can be seen from equations~(\ref{eq2.8}), this electron-electron interaction does not come from $\hat{\mathbf{A}}^{2}$ and can thus not be removed by the same redefinitions of frequencies and polarizations that would absorb $\hat{\mathbf{A}}^{2}$ in the velocity gauge. The dipole self-energy and the vector-potential-operator-squared terms are clearly not equivalent. 

We further perform a variable transform that effectively swaps between conjugate momentum and photon coordinate as~\cite{Ruggi PRA 2014}  
\begin{eqnarray}\label{eq2.10}
	i\frac{\partial}{\partial q_{\alpha} } \rightarrow p_{\alpha}\qquad \textrm{and}\qquad q_{\alpha} \rightarrow -i\frac{\partial}{\partial p_{\alpha}}.
\end{eqnarray}
The transformation above is merely a Fourier transformation of the mode $\alpha$ of the full wave-function 
\begin{equation}\label{eq2.11}
		\Psi'(..., q_{\alpha},...)\longrightarrow \Psi(...,p_{\alpha},...)=\frac{1}{\sqrt{2\pi}}\int^{\infty}_{-\infty} e^{- i q_{\alpha}p_{\alpha}}\Psi'(...,q_{\alpha},...)\textrm{d} q_{\alpha}.
\end{equation}
This variable transformation leaves the commutation relations unchanged. The Hamiltonian in length gauge is
\begin{eqnarray}\label{eq2.12}
	\hat{H}_L&=&-\frac{\hbar^2}{2m}\sum\limits^{N}_{i=1}\bi{\nabla}^2_i+\frac{1}{4\pi \epsilon_0} \sum\limits^{N}_{i< j}\frac{e^2}{|\bi{r}_i-\bi{r}_j|}+\sum\limits^{N}_{i=1}v_{ext}(\bi{r}_{i})\nonumber\\
	&+&\sum\limits^{M}_{\alpha=1}\left[ -\frac{\hbar\omega_{\alpha}}{2}\frac{\partial^2}{\partial p^2_{\alpha}}+\frac{\hbar\omega_{\alpha}}{2}\left(p_{\alpha}-\frac{Ce}{\hbar c}\frac{\bm{\epsilon}_{\alpha}\cdot \bi{R}}{\sqrt{\omega_{\alpha}}}\right)^2\right] .
\end{eqnarray}
This Hamiltonian contains the explicit bilinear electron-photon interaction 
\begin{equation}\label{eq2.14}
	\hat{V}_{int} = -\sum_{\alpha=1}^{M} \left(\bm{\lambda}_{\alpha}\cdot\bi{R}\right) p_{\alpha}\qquad \textrm{where}\qquad \bm{\lambda}_{\alpha}=\frac{\sqrt{\omega}_{\alpha}eC\bm{\epsilon}_{\alpha}}{c},
\end{equation}
as well as the dipole self-energy~\cite{Cohen, spohn2004, Ruggi PRA 2014}
\begin{equation}\label{eq2.15}
	\hat{\varepsilon}_{dip}  = \sum_{\alpha=1}^{M} \frac{\hbar\omega_{\alpha}}{2}\left( \frac{eC}{c\hbar} \frac{\bm{\epsilon}_{\alpha}\cdot\bi{R}}{\sqrt{\omega_{\alpha}}}\right)^{2}. 
\end{equation}
As we have seen from equations~(\ref{eq2.8}), these terms arise because the length-gauge transformation~(\ref{eq2.7}) mixes matter and photon degrees. Indeed, the coordinate $p_{\alpha}$ does not correspond anymore to a purely photonic quantity but rather to the electromagnetic displacement field that contains besides the electric field also the polarization of the matter system (see Sec.~\ref{subsec:Maxwell} for details). A different way to see that the length-gauge Hamiltonian mixes matter and photon degrees is how the simple translational invariance of the electronic subsystem is expressed in terms of the emerging polaritonic coordinates. \\

To do so we first take $v_{ext}(\bi{r})=0$, and thus the original Hamiltonian~(\ref{eq2.6}) becomes invariant under translations $\bi{r}\rightarrow \bi{r}+\bi{a}$, where $\bi{a}$ is an arbitrary vector. With the help of the translation operator on the configuration space of the electrons 
\begin{equation}
    \hat{T}(\bi{a})=\exp\left(\frac{i}{\hbar}\sum_{j=1}^{N}\bi{a}\cdot \hat{\bi{p}}_j\right)=\exp\left(\sum_{j=1}^{N} \bi{a}\cdot\nabla_j\right),
\end{equation}
this can be expressed equivalently by
\begin{equation}\label{eq2.21}
   [\hat{H}_{V},\hat{T}(\bi{a})]=0.
\end{equation}
Clearly, applying the same naive translation to the corresponding length-gauge Hamiltonian~(\ref{eq2.12}) will not work and $\hat{T}(\bi{a})$ will not commute with $\hat{H}_{L}$ in general. However, since the velocity gauge Hamiltonian is invariant under translations $\hat{H}_V = \hat{T}(\bi{a}) \hat{H}_V \hat{T}^{\dagger}(\bi{a})$ and by using the length-gauge transformation~(\ref{eq2.7}) we find that the length gauge Hamiltonian is invariant under
\begin{equation}\label{eq2.25}
    \hat{H}_{L}'=\hat{U}^{\dagger}\hat{T}(\bi{a})\hat{U}\hat{H}_{L}'\hat{U}^{\dagger}\hat{T}^{\dagger}(\bi{a})\hat{U}.
\end{equation}
Thus in the length gauge the translation operator is transformed as well via $\hat{T}_{L}'(\bi{a})=\hat{U}^{\dagger}\hat{T}(\bi{a})\hat{U}$, with the help of the Baker-Hausdorff-Campbell formula can be written as:
\begin{equation}\label{eq2.26a}
    \hat{T}_{L}'(\bi{a})=\hat{U}^{\dagger}\hat{T}(\bi{a})\hat{U} = \exp\left[\frac{i}{\hbar} \sum_{j=1}^{N}\bi{a}\cdot \left(\hat{\bi{p}}_j  + \frac{e}{ c}\hat{\mathbf{A}}\right)\right].
\end{equation}
Finally, after the Fourier transformation~(\ref{eq2.10}) we find 
\begin{eqnarray}\label{eq2.26b}
    \hat{T}_{L}(\bi{a})& =& \exp\left[\frac{i}{\hbar} \sum_{j=1}^{N}\bi{a}\cdot \left(\hat{\bi{p}}_j  + \frac{e}{ c}\sum^{M}_{\alpha=1}\frac{C \bm{\epsilon}_{\alpha}}{\sqrt{\omega_{\alpha}}}\left(-i \frac{\partial}{\partial p_{\alpha}}\right)\right)\right]\nonumber\\
    &=&\exp\left[\frac{i}{\hbar} \sum_{j=1}^{N}\bi{a}\cdot \hat{\bi{p}}_j + \frac{i}{\hbar}\sum^{M}_{\alpha=1}a_{\alpha}\left(-i\hbar\frac{\partial}{\partial p_{\alpha}}\right)\right],
\end{eqnarray}
where $a_{\alpha}= C e \bm{\epsilon}_{\alpha}\cdot N \bi{a}/\sqrt{\hbar^2 c^2 \omega_\alpha}$ . Thus the original translation of only the electronic subspace becomes a generalized translation in the full polaritonic configuration space of dimension $3N+M$ such that
\begin{equation}
 (\bi{r}_1,...,p_{1},...)\longrightarrow (\bi{r}_1+\bi{a},...,p_1+ C e \bm{\epsilon}_{1}\cdot N \bi{a}/\sqrt{\hbar ^2 c^2 \omega_1},...).
\end{equation}
That this is true can be easily verified if we use this polaritonic translation where both matter and photons are shifted and apply it to the length-gauge Hamiltonian~(\ref{eq2.12}). Indeed we find that the crucial term  
\begin{equation}
 \left(p_{\alpha}-\frac{Ce}{\hbar c}\frac{\bm{\epsilon}_{\alpha}\cdot \bi{R}}{\sqrt{\omega_{\alpha}}}\right) 
\end{equation}
is invariant, which implies that the naive picture that $\bi{r}$ corresponds to matter and $p$ corresponds to photon degrees is not valid. Both are mixtures of matter and photons and thus polaritonic in nature. We have therefore reformulated the original velocity-gauge Hamiltonian in a dressed matter-photon/polaritonic basis~\cite{grynberg2010, Johannes PNAS 2017, juzeliunas2002}.

Before we move on to investigate how neglecting the dipole self-energy term $\hat{\varepsilon}_{dip}$ affects non-relativistic QED in the long-wavelength limit, we point out a further important consequence of the dipole approximation. Upon assuming a uniform field, i.e., $e^{\pm i\bi{k}_n\cdot\bi{r}}\approx 1$, the energy expression~(\ref{eq2.3a}) does no longer hold in general. This problem already appears for a single photon mode, in which case the corresponding electric~(\ref{eq2.17}) and magnetic field~(\ref{eq2.18}) operators reduce to
\begin{eqnarray}
\hat{\textbf{E}} &=& C \frac{i}{c}\sqrt{\frac{\omega}{2}}\bm{\epsilon}\left[\hat{a} - \hat{a}^{\dagger} \right], \label{eq2.16a}\\
\hat{\textbf{B}} &=& C\frac{i}{c} \frac{\bi{k}\times\bm{\epsilon}}{\sqrt{2 \omega}}\left[\hat{a} - \hat{a}^{\dagger} \right]. \label{eq2.16b}
\end{eqnarray}
Using equations~(\ref{eq2.16a}) and (\ref{eq2.16b}) in equation~(\ref{eq2.3a}) results in
\begin{equation}\label{eq2.19}
\hat{H}_{p} = \hbar \omega \left( \hat{a}\hat{a}^{\dagger} + \hat{a}^{\dagger}\hat{a} - \hat{a}^{2} - \left(\hat{a}^{\dagger}\right)^{2}\right).
\end{equation}
Consequently we see that applying the dipole approximation to the original energy expression in terms of the electric and magnetic field operators will not lead to the usual form of the energy of the electromagnetic field. Hence it is crucial to only apply the dipole approximation after we have made the photon energy expression independent of the mode functions, i.e., Eq.~(\ref{eq2.2}). In Sec.~\ref{subsec:SemiClassics} we will see that also the semi-classical limit of light-matter systems depends on whether one first performs the dipole approximation and then the semi-classical limit or \textit{vice versa}.

\section{\label{sec:3}No ground-state without the dipole self-energy}

In the following we are interested in the spectral properties of the velocity and length gauge Hamiltonians, specifically in the question whether they support a ground-state. This question is paramount if we want to have a variational principle or employ an extension of ground-state density-functional theory~\cite{ruggenthaler2015ground} to treat non-relativistic QED in the dipole approximation. 

But before we want to make more precise what we mean by a ground-state and the dipole approximation. In most cases if the uncoupled, i.e., purely Coulombic, Hamiltonian has a ground-state then so does the minimal-coupling Hamiltonian~\cite{spohn2004}. A ground-state means that we cannot find any other state that has less energy, i.e., for all $\Psi$ in the self-adjoint domain of the Hamiltonian it holds that $\braket{\Psi}{\hat{H}\Psi}\geq E_0$, where $E_0$ is the ground-state energy. Indeed, for a broad class of potentials, e.g., $v_{ext}(\bi{r})\in L^2(\mathbb{R}^3)+L^{\infty}(\mathbb{R}^{3})$~\cite{spohn2004, Blanchard, Teschl} it holds that both Hamiltonians are bounded from below by some $E_0$ (but for such a broad class a ground-state does not necessarily exist, e.g., for $v_{ext}(\bi{r})=0$ we have only scattering states and the lower bound is $E_0 =0$). Thus in order to check for the existence of a ground-state we need to vary over all possible wave-functions in the domains of the respective Hamiltonians and show that they are bounded from below. This includes by construction also all functions that are non-zero only within a ball of finite radius, which can be located anywhere in $\mathbb{R}^3$, and infinitely many times differentiable (see as an example the function in equation~(\ref{eq3.10}))~\cite{spohn2004, Blanchard, Teschl}.

Since a ground-state is exponentially localized, i.e., falls off exponentially (but never becomes zero) away from the binding potential~\cite{spohn2004, Blanchard}, we can expect that if only modes with a wavelength much larger than the extension of the matter system are important, e.g., in the case of optical cavities, approximating the mode functions in equation~(\ref{eq2.4b}) by a constant will only slightly change the exponential tails of the ground-state wave-function. This is the working assumption of the dipole approximation of equation~(\ref{eq2.6}). The resulting velocity-gauge Hamiltonian is bounded from below and thus has the basic requirement for such an (exponentially localized) ground-state. Since the length-gauge Hamiltonian of equation~(\ref{eq2.12}) is merely a unitary transform, the same holds for $\hat{H}_{L}$. If the approximate Hamiltonian would be unbounded from below, then we cannot have an (exponentially localized) ground-state, contradicting the basic dipole assumption.

However, often the dipole self-energy of the electron $\hat{\epsilon}_{dip}$ that arises in the length-gauge picture is ignored based on the argument that it is a quantity which depends on the normalization volume of the field and for interactions of photons with an individual electron or atom one may take the limit $L^3\rightarrow \infty$ and in this case $\hat{\varepsilon}_{dip}\rightarrow 0$~\cite{Faisal}. The dipole self-energy is supposed to be important only in the thermodynamic limit where $N\rightarrow \infty$. Another reason why the dipole self-energy is often ignored in practice in fields like multi-photon processes in atomic and molecular physics~\cite{Faisal} or cavity and circuit QED~\cite{grynberg2010}, is due to simplified models of matter-photon interaction. For instance, assuming that only the ground-state $\ket{g}$ and the first excited state $\ket{e}$ of the bare matter system contribute to the dynamics such that $\hat{H}_e \sim E_g\ket{g}\bra{g} + E_e \ket{e}\bra{e}$ and that one mode described by $p = (\hat{a} + \hat{a}^{\dagger})/\sqrt{2}$ is in resonance with this transition $\hbar\omega = E_e-E_g$ we can approximate the length-gauged matter-photon Hamiltonian by~\cite{grynberg2010} 
\begin{equation}\label{eq3.3}
	\hat{H}_{R}' = \frac{\hbar \omega}{2}\hat{\sigma}_{z} + \hbar\omega\hat{a}^{\dagger}\hat{a}  - \frac{\hbar \Omega_R}{2}\hat{\sigma}_x(\hat{a}+\hat{a}^{\dagger}) + G \hat{\sigma}_x^{2} ,
\end{equation}
where we assumed that $\braket{g}{(\bm{\epsilon}\cdot\bi{R}) g} = \braket{e}{(\bm{\epsilon}\cdot\bi{R}) e} = \braket{g}{(\bm{\epsilon}\cdot\bi{R})^2 e} = 0 $, $\Omega_R=\sqrt{2 \omega} e C |\braket{g}{\bm{\epsilon}\cdot\bi{R} e}|/(c \hbar)$ and $G = \braket{g}{\hat{\epsilon}_{dip} g} + \braket{e}{\hat{\epsilon}_{dip} e}$. Here the last term $ \hat{\sigma}_x^2$ is just the identity operator in the two-dimensional electronic subspace and consequently just gives a constant energy offset. 
Thus, for this reduced model we can drop the last term of the Hamiltonian of equation~(\ref{eq3.3}), which corresponds to the dipole self-energy, and we obtain the Rabi model Hamiltonian~\cite{grynberg2010, Braak}
\begin{equation}\label{eq3.4}
\hat{H}_{R} = \frac{\hbar \omega}{2}\hat{\sigma}_{z} + \hbar\omega\hat{a}^{\dagger}\hat{a}  - \frac{\hbar \Omega_R}{2}\hat{\sigma}_x(\hat{a}+\hat{a}^{\dagger}).
\end{equation}
In this Rabi Hamiltonian $\hat{H}_{R}$ we can perform the rotating wave approximation (RWA)~\cite{grynberg2010, Shore JMO 1993} in which $\hat{\sigma}_x(\hat{a}+\hat{a}^{\dagger}) \rightarrow \hat{\sigma}_{+}\hat{a}+\hat{\sigma}_{-}\hat{a}^{\dagger}$, where $\hat{\sigma}_x = \hat{\sigma}_{+} + \hat{\sigma}_{-}$, and we obtain the Jaynes-Cummings model Hamiltonian~\cite{grynberg2010, Shore JMO 1993}
\begin{equation}
    \hat{H}_{JC} = \frac{\hbar \omega}{2}\hat{\sigma}_{z} + \hbar\omega\hat{a}^{\dagger}\hat{a}  - \frac{\hbar \Omega_R}{2}\left(\hat{\sigma}_{+}\hat{a}+\hat{\sigma}_{-}\hat{a}^{\dagger}\right).
\end{equation}
In the case that we consider $N$ identical matter systems coupled via the RWA we obtain the Dicke model Hamiltonian~\cite{Garraway}
\begin{equation}
    \hat{H}_{D}=\hbar\omega\hat{a}^{\dagger}\hat{a} -\frac{\hbar \Omega_R}{2}\left(\hat{S}_{+}\hat{a}+\hat{S}_{-}\hat{a}^{\dagger}\right)+ \frac{\hbar \omega}{2}\hat{S}_{z}
\end{equation}
where $\hat{S}_{+}\;,\hat{S}_{-}$ and $\hat{S}_{z}$ are collective operators defined as 
\begin{equation}
    \hat{S}_{\pm}=\sum\limits^N_{i=1}\hat{\sigma}^{i}_{\pm} \qquad \textrm{and}\qquad \hat{S}_{z}=\sum\limits^{N}_{i=1}\hat{\sigma}^{i}_{z}.
\end{equation}
Thus more advanced models that are based on the Rabi or Jaynes-Cummings Hamiltonian, e.g., the Dicke model, do often not contain the dipole self-energy term.

\begin{figure}
\includegraphics[width=3in, height=2.5in]{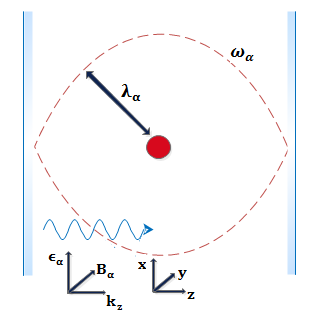}
\caption{\label{fig:one} Schematic illustration of a matter system in a cavity coupled to a single-mode field, $\bm{\lambda}_{\alpha}$ is the coupling-strength polarization vector of the mode $\alpha$ as defined in equation~(\ref{eq2.14}). The propagation is along the $z$-direction and the field is polarized perpendicularly. The extension of the matter system (in red) is assumed to be much smaller such that we can effectively assume the electron to be in $\mathbb{R}^{3}$.}
\end{figure}

Now, to investigate how the dipole self-energy term impacts the spectral properties of the length-gauge Hamiltonian we will consider what happens upon ignoring this harmonic self-interaction. For simplicity we will restrict to the one electron and one photon mode case, i.e., $N=M=1$. The general case of arbitrary many electrons and modes can be treated in a similar manner and is presented for completeness in Appendix~A. In the simple case the Hamiltonian~(\ref{eq2.12}) takes the  form
\begin{eqnarray}\label{eq3.5}
	\hat{H}_{L}=-\frac{\hbar}{2m}\bi{\nabla}^2 -\frac{\hbar\omega}{2}\frac{\partial^2}{\partial p^2} +\frac{\hbar\omega}{2}\left(p-\frac{Ce}{\hbar c}\frac{\bm{\epsilon}\cdot\bi{r}}{\sqrt{\omega}}\right)^2+v_{ext}(\bi{r}).
\end{eqnarray}
It describes a single-active electron system coupled to a single mode of a high-Q cavity, i.e., we do not consider dissipation of the photon mode. The inclusion of dissipation, however, will not change the outcome of the following discussion. As is shown in Fig.~\ref{fig:one}, the propagation direction of the photons can be assumed along the $z$-direction and the field is polarized perpendicular to it. We further assume that the electron can leave the cavity also through the mirrors and thus we consider the full space $\mathbb{R}^3$ in accordance to the minimal-coupling and the uncoupled problem. We will comment on this assumption at the end of this section. The Hamiltonian without the dipole self-energy $\hat{H}^{'}=\hat{H}_{L}-\hat{\varepsilon}_{dip}$ reads as
\begin{equation}\label{eq3.7}
	\hat{H}^{'}=-\frac{\hbar}{2m}\bi{\nabla}^2 -\frac{\hbar\omega}{2}\frac{\partial^2}{\partial p^2} +\frac{\hbar\omega}{2}p^2-\left(\bm{\lambda}\cdot\bi{r}\right) p+v_{ext}(\bi{r}),
\end{equation}
where the photon polarization is included in
\begin{equation}
    \bm{\lambda}=\frac{Ce\sqrt{\omega}\bm{\epsilon}}{c}.
\end{equation}
The Hamiltonian $\hat{H}_{L}$, as discussed above, is bounded from below. However, is $\hat{H}'$ bounded from below as well, which is a necessary prerequisite for a ground-state to exist? To answer this question, let us consider a trial wave-function and we calculate its energy with respect to $\hat{H}^{'}$. The photonic part of the wave-function is described by the function
\begin{equation}\label{eq3.8}
	\Phi(p)=\frac{1}{\sqrt{2}}\left[\phi_1(p)+\phi_{2}(p)\right],
\end{equation}
where the functions $\phi_1(p)$ and $\phi_{2}(p)$ are the normalized ground and first excited eigen-states of the harmonic oscillator~\cite{Griffiths, Teschl}. For the electronic part of the wave-function we consider
\begin{eqnarray}\label{eq3.10}
	F_{a}(\bi{r})=
	\begin{cases}
		\mathcal{N}\exp[-\frac{1}{1-|\bi{r}-\bi{a}|^2}],\qquad \textrm{if} \qquad|\bi{r}-\bi{a}|<1 \\
		 \qquad0, \hspace{2,8cm} \textrm{if}\qquad|\bi{r}-\bi{a}|\geq 1 \\
		\textrm{where} \qquad\bi{a}=a\bm{\kappa}, \qquad a\in \mathbb{R}
	\end{cases}
\end{eqnarray}
\begin{figure}
\includegraphics[width=3in, height=2.5in]{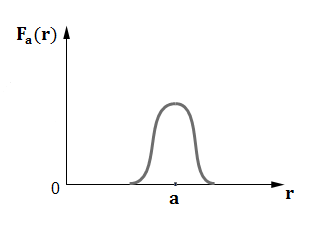}
\caption{\label{fig:two}Schematic one-dimensional illustration of the mollifier $F_a(\bi{r})$. The mollifier $F_a(\bi{r})$ and also all its derivatives are non-zero only within the unit ball $|\bi{r}-\bi{a}|<1$.}
\end{figure}
where $\bm{\kappa}$ is an arbitrary vector and $a$ an arbitrary parameter. This function, shown in Fig.~\ref{fig:two}, is non-zero only within the unit ball $|\bi{r}-\bi{a}|<1$, is normalized, with $\mathcal{N}$ being its normalization constant, and is infinitely many times differentiable. We could have chosen any other well-behaved function but we fix this one for definiteness. Thus, the complete wave-function will be the tensor product of $F_{a}(\bi{r})$ and $\Phi(p)$,
\begin{equation}\label{eq3.11}
	\Psi=F_{a}(\bi{r})\otimes\Phi(p).
\end{equation}
The wave-function $\Psi$ is part of the domain of $\hat{H}_L$ since $\braket{\hat{H}_L \Psi}{\hat{H}_L\Psi} < \infty$ and also of $\hat{H}'$. The energy with respect to $\hat{H}'$ consists of four different terms


\begin{eqnarray}\label{eq3.13}
\braket{\Psi}{\hat{H}^{'}\Psi}= -\frac{\hbar^2}{2m}\braket{F_{a}}{\bi{\nabla}^2 F_{a}}+\braket{\Phi}{\hat{H}_p \Phi}+\braket{F_{a}}{\hat{V}_{ext} F_{a}}+ \braket{\Psi}{\hat{V}_{int}\Psi}.
\end{eqnarray}
The first term is the kinetic energy of the electron, the second is the energy of the photons, the third is the potential energy and the last term is the contribution of the bilinear interaction to the energy. The kinetic energy of the electron reads
\begin{eqnarray}\label{eq3.14}
	-\frac{\hbar^2}{2m}\braket{F_{a}}{\bi{\nabla}^2 F_{a}}=-\frac{\hbar^2|\mathcal{N}|^2}{2m}\int\limits_{|\bi{r}-\bi{a}|<1} e^{-\frac{1}{1-|\bi{r}-\bi{a}|^2}} \bi{\nabla}^2 \left(e^{-\frac{1}{1-|\bi{r}-\bi{a}|^2}}\right)\d^3r.
\end{eqnarray}
Since the kinetic energy operator is translational invariant, we can perform the transformation $\bi{r} \rightarrow \bi{r}+\bi{a}$ without changing its value. Thus we have
\begin{eqnarray}\label{eq3.17}
-\frac{\hbar^2}{2m}\braket{F_{a}}{\bi{\nabla}^2 F_{a}}&=&-\frac{\hbar^2}{2m}\braket{F_{0}}{\bi{\nabla}^2 F_{0}}=\\
&=&-\frac{\hbar^2|\mathcal{N}|^2}{2m}\int\limits_{|\bi{r}|<1} e^{-\frac{1}{1-|\bi{r}|^2}} \bi{\nabla}^2 \left(e^{-\frac{1}{1-|\bi{r}|^2}}\right)\d^3r=T<\infty.
\end{eqnarray}
The second term becomes
\begin{eqnarray}\label{eq3.18}
	\braket{\Phi}{\hat{H}_p \Phi}=\frac{1}{2}(E_1+E_2) \qquad\textrm{where}\qquad E_n=\hbar\omega\,(n+\frac{1}{2})\qquad \forall \; n \in \mathbb{N}\;.
\end{eqnarray}
Next we estimate the potential energy of the electron due to the external potential. Without loss of generality (we comment on this at the end of this section) we choose the external potential to be some negative, attractive potential and consequently the contribution of $v_{ext}$ will be negative as is the usual case for atomic and molecular systems, i.e.,
\begin{equation}\label{eq3.19}
	\braket{F_{a}}{\hat{V}_{ext} F_{a}}=-V_{a} \qquad \textrm{where}\qquad V_{a}\geq0.
\end{equation}
Finally, the energy of the bilinear interaction
\begin{eqnarray}\label{eq3.20}
\braket{\Psi}{\hat{V}_{int} \Psi}=-\braket{\Phi}{p \Phi} \braket{F_{a}}{\left( \bm{\lambda}\cdot \bi{r}\right) F_{a}} 
\end{eqnarray}
with $\braket{\Phi}{p \Phi}=1$ becomes 
\begin{equation}
    \braket{\Psi}{\hat{V}_{int} \Psi}=-\bm{\lambda}\cdot\braket{F_{a}}{\bi{r}F_{a}}=-|\mathcal{N}|^2\int \limits_{|\bi{r}-\bi{a}|<1}\bm{\lambda}\cdot \bi{r}e^{-\frac{2}{1-|\bi{r}-\bi{a}|^2}} \d^3r.
\end{equation}
We perform again the translation $\bi{r}\rightarrow \bi{r}+\bi{a}$ and the integral will change as follows
\begin{eqnarray}\label{eq3.21}
 \braket{\Psi}{\hat{V}_{int} \Psi} = -|\mathcal{N}|^2\int \limits_{|\bi{r}|<1}\bm{\lambda}\cdot \bi{r}\;e^{-\frac{2}{1-|\bi{r}|^2}}d^3r -\bm{\lambda}\cdot\bi{a}\;\braket{F_0}{F_0}=-\bm{\lambda}\cdot\bi{a}=-a 
\end{eqnarray}
where we now have chosen $\bm{\kappa} = \bm{\lambda}/|\bm{\lambda}|^2$. The first integral of equation (\ref{eq3.21}) is zero. From the expression above we see that the contribution of the bilinear interaction to the energy is proportional to $-a$. Summing all the different contributions, we find the total energy to be
\begin{equation}\label{eq3.22}
	\braket{\Psi}{\hat{H}^{'}\Psi}= T+\frac{1}{2}(E_1+E_2)-V_a-a \leq T+\frac{1}{2}(E_1+E_2)- a  \sim -a\;.
\end{equation}
From this result it becomes clear that the Hamiltonian is unbounded from below, since the parameter $a$ can be chosen arbitrarily ($F_{a}$ can be moved further and further away from the origin) and we can therefore lower the energy of $\hat{H}^{'}$ as much as we want. Thus, we conclude that no ground-state exists without the dipole self-energy. This is not so surprising because we have subtracted a harmonic oscillator from the Hamiltonian~(\ref{eq3.5}) and despite claims in literature this term is dominant and cannot be discarded. Maybe even more striking is that if we transform $\hat{H}'$ back into the original velocity gauge, we find due to the fact that $\hat{\epsilon}_{dip}$ commutes with the transformations
\begin{eqnarray}
  \hat{H}_V'=\frac{1}{2m} \left[i \hbar \mathbf{\nabla} +\frac{e}{c} \hat{\mathbf{A}}\right]^2 +  v_{ext}(\bi{r}) 
	+\hbar \omega \left(\hat{a}^{\dagger} \hat{a} + \frac{1}{2}  \right) -  \frac{1}{2 \hbar}\left(\bm{\lambda} \cdot \bi{r} \right)^2.
\end{eqnarray}
It seems obvious that discarding the dipole self-energy will make the system unstable. Let us further point out that $\hat{H}^{'}$ will in general not have any eigen-states and we will have a purely continuous spectrum with only scattering states. Thus to use $\hat{H}^{'}$ makes only sense in a time-dependent setting. This difference we will also encounter in Sec.~\ref{subsec:SemiClassics} in the case of the semi-classical limit.

Let us finally comment on the choice of considering the particles to be in full space while we used a box for the construction of the photon field, as well as a purely negative external potential $v_{ext}(\bi{r})$. It is, first of all, straightforward to allow for different lengths of the quantization volume in different directions, such that we can model besides free space also a planar cavity. Further, it would also not be a problem to use different boundary conditions, say, zero boundary conditions along the $z$-direction (see Fig.~\ref{fig:one}) and periodic ones along $x$ and $y$ and finally take the limit to infinity for these unconstrained directions~\cite{spohn2004}. But since all these considerations become superfluous for the general investigation about the spectral properties of the length-gauge Hamiltonian with or without the dipole self-energy we employed the periodic setting. The explicit form of the photon field does not change the harmonic nature of the dipole self-energy. For investigating the ground-state of the length-gauge Hamiltonian it does, however, matter if we enclosed the particles in a finite volume. In this case we could not lower the energy of $\hat{H}^{'}$ indefinitely. But we would find a ground-state that is localized at the edge of the box provided the box is large enough. Such a wave-function is not a physical ground-state and we would have a maximally allowed box length for a given atomic or molecular Hamiltonian before the subtracted dipole self-energy becomes dominant. The inclusion of the dipole self-energy, however, allows for a box-size independent limit, which seems a physically desirable property. Furthermore, if we consider the case of a high-Q cavity (see Fig.~\ref{fig:one}), it is reasonable to only constrain the particles along the quantization direction. Since the field is perpendicular, the particles will be ionized along the unconstrained directions and again we are unbounded from below. Thus to use the full space $\mathbb{R}^3$ is very sensible and we could even model the barrier of the mirrors for the particles by just putting a very large repulsive potential at the assumed positions of the mirrors.
But even a very large repulsive, i.e., positive, potential will not help to guarantee a ground-state. Since all potentials we consider are in the Banach space $L^{2}(\mathbb{R}^3)+L^{\infty}(\mathbb{R}^3)$, in the limit $|\bi{r}| \rightarrow \infty$ only the bounded part of the potential survives. Thus by shifting $F_{a}$ further and further away from the origin only an energy contribution proportional to the limiting constant $v_{ext}(\bi{r}) \rightarrow v_{ext}^{\infty}$ contributes. Thus also a positive potential in $L^{2}(\mathbb{R}^3)+L^{\infty}(\mathbb{R}^3)$ will not be able to compensate the linear decrease in energy due to the coupling.

\section{\label{sec:4}Significance of the dipole self-energy}

Clearly, as shown above the dipole self-energy is very important and neglecting this term dramatically changes the physical properties of the combined light-matter system. Its harmonic confinement is an explicit manifestation that we assumed the wavelength of the photon field to be much larger than the extend of the matter system. Without it an exponentially localized ground-state becomes impossible. Even for arbitrarily small but finite coupling to the field the system decays and has no stable ground-state. The harmonic self-energy term is therefore necessary to guarantee a variational principle and to allow for static properties of the combined matter-photon system. But besides this dramatic effect, there are several other points where the dipole self-energy becomes important.

\subsection{The Maxwell equations in matter}
\label{subsec:Maxwell}

Let us first consider the consequences of the dipole self-energy for the photon field. As previously stated, a peculiarity of the length gauge transformation is the mixture of the electronic and photonic degrees of freedom. Therefore the ``photonic'' coordinates $p_{\alpha}$ do not correspond to pure electromagnetic quantities. This can be seen if we determine how the electric-field $\hat{\bi{E}}=-\frac{1}{c}\frac{\d \hat{\bi{A}}}{\d t}$ looks like in the length-gauge picture. From the definition of the vector-potential operator~(\ref{eq2.5}) in the length-gauge picture, i.e., 
\begin{equation}\label{eq4.10}
\hat{\bi{A}} = \sum_{\alpha=1}^{M} \hat{\bi{A}}_{\alpha}, \qquad \textrm{where} \qquad \hat{\bi{A}}_{\alpha} = - \frac{iC\bm{\epsilon}_{\alpha}}{\sqrt{\omega_{\alpha}}}\frac{\partial}{\partial p_{\alpha}},
\end{equation}
the electric field becomes  
\begin{equation}\label{eq4.11}
\hat{\bi{E}}=-\frac{i}{c\hbar}[\hat{H}_{L},\hat{\bi{A}}]=\sum_{\alpha=1}^{M} \hat{\bi{E}}_{\alpha} \qquad \textrm{where} \qquad \hat{\bi{E}}_{\alpha} = \frac{C\bm{\epsilon}_{\alpha}}{c}\sqrt{\omega_{\alpha}}\left(p_{\alpha}-\frac{Ce\bm{\epsilon}_{\alpha}\cdot \bi{R}}{\hbar c\sqrt{\omega}_{\alpha}}\right).
\end{equation}
By defining the polarization operator 
\begin{equation}\label{eq4.12}
\hat{\bi{P}}=\epsilon_0\sum_{\alpha=1}^{M}\bm{\epsilon}_{\alpha} \frac{C^2e \left(\bm{\epsilon}_{\alpha}\cdot \bi{R}\right)}{\hbar c^2},
\end{equation}
we find that the $p_{\alpha}$ actually correspond to the displacement field operator
\begin{equation}\label{eq4.13}
\hat{\bi{D}}=\epsilon_0\sum_{\alpha=1}^{M}\frac{C\sqrt{\omega_{\alpha}}\bm{\epsilon}_{\alpha}}{c}p_{\alpha}
\end{equation}
of the Maxwell equations in matter, i.e., $\hat{\bi{D}}= \epsilon_0\hat{\bi{E}}+ \hat{\bi{P}}$. This result, however, would still hold even if we discarded the dipole self-energy term. The omittance of the dipole self-energy term will show up in higher moments and will lead to a violation of the equations of motion.

To see this we first consider equations of motion of $\hat{\bi{A}}$ and $\hat{\bi{E}}$ for the case of the Hamiltonian with the dipole self-energy. By taking the first time-derivative for the electric field, i.e., $\frac{\d^2}{\d t^2} \hat{\bi{A}} = -c \frac{\d}{\d t} \hat{\bi{E}}$, we obtain
\begin{equation}\label{eq4.14}
\frac{\d^2}{\d t^2}\hat{\bi{A}} + \sum_{\alpha}\omega_{\alpha}^2  \hat{\bm{A}}_{\alpha} =i\sum_{\alpha =1}^{M}\bm{\epsilon}_{\alpha}\;\frac{C^2e}{mc}\sum^{N}_{i=1}(\bm{\epsilon}_{\alpha}\cdot\nabla_{i}) .
\end{equation}
The equation of motion is the mode resolved inhomogeneous Maxwell equation, where the inhomogeneity corresponds to the projections on the transversal part of the current operator $\hat{\bi{J}}(\bi{r})=\frac{e \hbar}{2 m i}\sum_{i=1}^N \{\delta^3(\bi{r}-\bi{r}_i) \overrightarrow{\nabla}_i - \overleftarrow{\nabla}_i \delta^3(\bi{r}-\bi{r}_i)   \}$. It is important to note that this is indeed the full physical current since due to the length-gauge transformation this also contains the diamagnetic part due to the vector-potential operator~\cite{stefanucci2013}. For the equation of the electric field we make the specific choice $v_{ext}(\bi{r}) = 0$, since this simplifies the further analysis. This choice also includes the special but important case of the homogeneous electron gas. The Maxwell equation for the electric field then becomes 
\begin{equation}\label{eq4.15}
\frac{\d^2}{\d t^2} \hat{\bi{E}}=-\sum_{\alpha=1}^{M}\omega^2_{\alpha}\hat{\bi{E}}_{\alpha}-\frac{C^{2}e^{2} N}{m\hbar c^2}\hat{\bi{E}}=-\sum_{\alpha=1}^{M}(\omega^2_{\alpha}+\omega^2_p)\hat{\bi{E}}_{\alpha}.
\end{equation}
The equation of motion of the electric field follows the well-known mode resolved Maxwell equation. Moreover, we find that there are two contributions to the frequencies of the electric field. The frequencies $\omega_{\alpha}$, which are the bare frequencies of the photons, and the matter contribution $\omega_p=\left(\frac{C^2e^{2} N}{m\hbar c^2}\right)^{\frac{1}{2}}$. After rearranging the constants the matter contribution turns out to be the plasma frequency
\begin{equation}\label{eq4.16}
\omega^2_p=\frac{ne^2}{m\epsilon_0 },
\end{equation}
where $n=N/L^3$. The total frequencies of the electric field therefore are
\begin{equation}\label{eq4.17}
\widetilde{\omega}^{2}_{\alpha} = \omega^{2}_{\alpha} + \omega_{p}^{2}. 
\end{equation}
This change in frequency is known as the "depolarization shift"~\cite{Ando RMP 1982} and has been observed experimentally, for instance, in resonant matter-photon systems in the ultra-strong coupling regime~\cite{Bogoliubov proced., Todorov PRL 2010,Todorov PRB 2012,Todorov PRX 2014,Vasanelli CRP 2016}. On the other hand, if we do ignore the dipole self-energy and make the specific choice of $v_{ext}(\bi{r}) = 0$, the equation for the electric field becomes
\begin{equation}\label{eq4.18}
\frac{\d^2}{\d t^2}\hat{\mathbf{E}}+\sum_{\alpha=1}^{N}\omega^2_{\alpha}\hat{\mathbf{E}}_{\alpha}=-\frac{C^2e^2N}{m\hbar c^2}\hat{\mathbf{D}}.
\end{equation}
The omittance of the dipole self-energy therefore leads to a wrong description of the electromagnetic part of the coupled system since in the right-hand side of equation~(\ref{eq4.18}) we do not have the electric field $\hat{\bi{E}}$ but the displacement field $\hat{\bi{D}}$. The reason is the mixing of matter and photon degrees in the length-gauge transformation. Therefore, the dipole self-energy term must not be ignored in order to get a complete physical description.


\subsection{The semi-classical limit}
\label{subsec:SemiClassics}

In the length gauge Hamiltonian~(\ref{eq2.12}), we showed that the dipole self-energy term $\hat{\varepsilon}_{dip}$ emerges due to the photon coordinates. This term will therefore not appear if we perform the semi-classical limit already in the original velocity-gauge Hamiltonian. Hence, the semi-classical limit performed \textit{after} the length-gauge transformations will be different. This is a further point where the dipole self-energy term becomes significant.

The standard semi-classical limit treats the vector potential in equation~(\ref{eq2.6}) as an external field that interacts with the electronic system. Since a uniform, time-independent classical vector potential has no physical effect, one needs to go to a time-dependent vector potential $\bi{A}(t)$ which gives rise to an electric field via $\bi{E}(t) = -\frac{1}{c}\frac{\d}{\d t} \bi{A}(t)$. Performing then the length-gauge transformation~(\ref{eq2.7}), where we now use a time-dependent classical vector potential, will eliminate the $\bi{A}^2$ and the $\bi{A}\cdot\nabla$ terms as in the quantum case. However, the new interaction between field and matter now emerges with $\Psi_{L}(t) = \hat{U}^{\dagger}(t)\Psi_{V}(t)$ due to 
\begin{equation}
 i \hbar \frac{\partial}{\partial t} \Psi_{L}(t) = - e \bi{E}(t)\cdot \bi{R} \Psi_{L}(t) + \hat{U}^{\dagger}(t) i \hbar \frac{\partial}{\partial t} \Psi_{V}(t),
\end{equation}
not due to the transformation of the photon coordinates as shown in equation~(\ref{eq2.8}). Consequently the standard semi-classical length-gauge Hamiltonian
\begin{equation}\label{eq4.3}
\hat{H}_{sc}(t)=\hat{T}_e + \hat{W}_{e} + \hat{V}_{ext} - e\mathbf{R}\cdot \mathbf{E}(t)
\end{equation}
is based on the time-dependent Schr\"odinger equation. If we now take a static electric field and turn this into an eigen-value problem we encounter the same problems as with $\hat{H}^{'}$. We will have no ground-state unless $\bi{E} = 0$. Since no ground-state exists, this raises problems, e.g., if we want to treat the Stark effect non-perturbatively. The standard semi-classical limit for $\bi{E} \neq  0$ only makes sense as a time-dependent problem.

If we now perform the semi-classical limit \textit{after} the length-gauge transformation, i.e., we start from the Hamiltonian~(\ref{eq2.12}), discard the ``kinetic'' term of the modes and treat $p_{\alpha}$ as a number, we get a coupled Maxwell-Schr\"odinger equation
\begin{eqnarray}\label{eq4.4}
	\hat{H}_{asc}'=\hat{T}_e + \hat{W}_{e} + \hat{V}_{ext} + \sum_{\alpha = 1}^{M}\frac{\hbar\omega_{\alpha}}{2}\left(p_{\alpha}-\frac{Ce}{\hbar c}\frac{\bm{\epsilon}_{\alpha}\cdot \bi{R}}{\sqrt{\omega_{\alpha}}}\right)^2,
\end{eqnarray}
where we determine $p_{\alpha}$ from its equation of motion using the Hamiltonian~(\ref{eq2.12})
\begin{equation}
 \frac{\d^2}{\d t^2} p_{\alpha}(t) = - \omega_{\alpha}^2 p_{\alpha}(t) + \frac{\omega_{\alpha}}{\hbar}\bm{\lambda}_{\alpha}\cdot\bi{R}(t).
\end{equation}
Here $\bi{R}(t)= \braket{\psi(t)}{\bi{R} \psi(t)}$ and $\psi(t)$ a purely electronic wave-function. For a static problem we therefore find, in accordance to the fact that $p_{\alpha}$ actually corresponds to the displacement field, that we are left with the polarization term only, i.e., in equilibrium $\bi{D} = \bi{P}[\psi] = \braket{\psi}{\hat{P} \psi}$ and the electric field is zero as it should for an eigen-state~\cite{ruggenthaler2015ground, dimitrov2017}. If we further choose a symmetric binding potential at the origin, the eigen-states have zero dipole-moment and we reduce to the usual Hamiltonian $\hat{H}_e$ plus the dipole-self energy. If we want to have a non-zero electric field we have to couple the photons to a time-dependent external current~\cite{tokatly-2013, Ruggi PRA 2014} that gives rise to the same $\bi{E}(t)$ as in the standard semi-classical limit and if we make this field static we arrive (up to a gauge shift) at an alternate form of the semi-classical limit
\begin{eqnarray}\label{eq:dynamicalmeanfieldnl}
 \hat{H}_{asc} = \hat{T}_e + \hat{W}_{e} + \hat{V}_{ext} - e \bi{R}\cdot \bi{E} -\frac{e}{\epsilon_0} \bi{R}\cdot \bi{P}[\psi] + \hat{\varepsilon}_{dip},
\end{eqnarray}
which is a non-linear equation for the electronic subsystem only. It is bounded from below due to the dipole self-energy term and thus supports ground-states in contrast to the standard limit. If we instead fix the displacement $\bi{D}$, which implies that we \textit{apriori} know the total field consisting of $\bi{E}$ and the induced field $\bi{P}$, i.e., we just fix $p_{\alpha}$ in equation~(\ref{eq4.4}) without coupling to the Maxwell equation, we can remove the non-linearity and find
\begin{eqnarray}\label{eq:dynamicalmeanfield}
 \hat{H}_{asc}' = \hat{T}_e + \hat{W}_{e} + \hat{V}_{ext} - \frac{e}{\epsilon_0} \bi{R}\cdot \bi{D} + \hat{\varepsilon}_{dip}.
\end{eqnarray}
Indeed, if we find an eigen-state $\psi$ of equation~(\ref{eq:dynamicalmeanfieldnl}) for a given $\bi{E}$ and we determine the polarization $\bi{P}[\psi]$, we can use the resulting $\bi{D}$ in equation~(\ref{eq:dynamicalmeanfield}) and trivially recover the same solution $\psi$.

 Finally, by using $\hat{\bi{A}}_{tot}(t) = (\hat{\bi{A}} + \bi{A}(t))$ and combining both above derivations we can include also time-dependent fields. Note, however, that we can always exchange an external field by an appropriately chosen external time-dependent current~\cite{dimitrov2017}. To conclude, the semi-classical limit performed \textit{after} the length-gauge transformations supports eigen-states due to the presence of the dipole self-energy term in contrast to the standard semi-classical limit. It therefore allows to treat equilibrium effects, such as the Stark effect, non-perturbatively but can also be applied to non-equilibrium situations, such as the ac-Stark shift.

\section{Conclusion and Outlook}

In this work we have shown the fundamental role played by the usually neglected dipole self-energy term for coupled light-matter systems in the long-wavelength limit. Without it no ground-state of the combined photon-matter system can exist. This dipole self-energy term is \textit{not equivalent} to the vector-potential operator squared and can therefore not be absorbed by simple redefinitions of frequency and polarization. Besides this key result we have highlighted several important consequences of the use of the length-gauge form of non-relativistic QED, such as the emergence of polaritonic coordinates and the corresponding change of the translational invariance. Moreover, we have shown how the dipole self-energy influences the corresponding Maxwell equations and leads, e.g., to the ``depolarization shift'' in the ultra-strong coupling regime. Further, we have introduced a different semi-classical limit where the dipole self-energy term appears which makes this semi-classical Hamiltonian bounded from below, in contrast to the standard form, and allows to treat, e.g., the Stark shift non-perturbatively.

The basic assumption we employed to establish that no ground-state exists without the dipole self-energy is that we consider the particles in infinite space. In contrast, the dipole approximation \emph{with} the dipole self-energy is bounded from below for a broad class of potentials and thus has the prerequisite to support (exponentially localized) ground-states, which is the working assumption of the dipole approximation. We have argued in the end of Sec.~\ref{sec:3} why it is physically reasonable similarly to the full minimal-coupling and uncoupled (purely Coulombic) problem to consider the particles in infinite space. We note that we did not consider under which conditions the dipole approximation is reasonable in the first place. Although it is expected that in most cases it will have an (exponentially localized) ground-state if also the corresponding minimal-coupling problem has a ground-state, it is not clear a priori whether only the exponential tails of the ground-state wave-function will be affected by the approximation. This will be subject of future research.

The presented results have a direct consequence for static properties of coupled light-matter systems in the long-wavelength limit, both for the case of quantized electromagnetic fields as well as for static classical electric fields. The description of such properties beyond model-system or perturbation approaches needs to include the dipole self-energy. \textit{Ab-initio} techniques for coupled light-matter systems, such as ground-state density-functional theory~\cite{dreizler-gross} and its extension to quantized fields~\cite{ruggenthaler2015ground}, therefore crucially depend on this harmonic term. As shown in, e.g., Ref.~\cite{Johannes PNAS 2017} this term dramatically changes how the properties of a matter-photon system adopt upon increasing the effective coupling strength. Since changing chemical properties due to strong matter-light coupling in, e.g., polaritonic chemistry~\cite{ebbesen-2016} is experimentally feasible, reliable \textit{ab-initio} methods that can handle such situations become invaluable. But the inclusion of the dipole self-energy term will clearly also have an impact on the time-dependent situation. Since the spectrum of the coupled light-matter Hamiltonian with and without this term is dramatically different, the resulting time-dependent wave functions will differ. Under which conditions these differences become evident is an interesting future research perspective. Furthermore, the dipole self-energy influences also the photonic observables. The presented result can therefore shed new light on, e.g., the debate of superradiance in Dicke-type models~\cite{Vukics PRL, Vukics}. By comparing, e.g., the ground-state energy of the Dicke model~\cite{bastarrachea2011} to  \textit{ab-initio} calculations when changing the coupling strength, one can test the reliability of the standard Dicke model in capturing the correct physical behaviour.  

 \begin{acknowledgments}
We acknowledge financial support from the European Research Council (ERC-2015-AdG-694097), Grupos Consolidados (IT578-13) and European Union Horizon 2020 program under Grant Agreement 676580 (NOMAD).
\end{acknowledgments}
 
\appendix

\section{Many-body Hamiltonian without dipole self-energy}

\begin{figure}
\includegraphics[width=3in, height=2.5in]{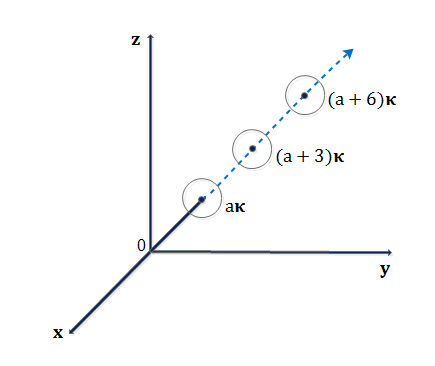}
\caption{\label{fig:three}Graphic representation of the localization of the electronic wave-function $\Psi_e$. The mollifiers are put on on an equally spaced grid along the vector $\bm{\kappa}$, e.g, such that there is no overlap between them.}
\end{figure}

In this appendix we consider the general case of non-relativistic QED in the long-wavelength limit for $N$ electrons interacting with $M$ photon modes described by equation~(\ref{eq2.12}). In order to investigate the impact of the dipole self-energy we drop this term, leading to the Hamiltonian
\begin{eqnarray}\label{eqa.5}
\hat{H}^{'}&=&\hat{H}_{L}-\hat{\varepsilon}_{dip}=-\frac{\hbar^2}{2m}\sum^{N}_{i=1}\bi{\nabla}^2_i+\frac{1}{4\pi\epsilon_0}\sum^{N}_{i< j}\frac{e^2}{|\bi{r}_i-\bi{r}_j|}+\sum^{N}_{i=1}v_{ext}(\bi{r}_{i})\nonumber\\
&+&\sum^{M}_{\alpha=1}\left[ -\frac{\hbar\omega_{\alpha}}{2}\frac{\partial^2}{\partial p^2_{\alpha}}+\frac{\hbar\omega_{\alpha}}{2}p^2_{\alpha}-\left(\bm{\lambda}_{\alpha}\cdot\bi{R}\right) p_{\alpha}\right].
\end{eqnarray}
We consider the energy of a trial wave-function with respect to $\hat{H}'$ and show that we can lower the energy indefinitely. For the electronic part of the wave-function we use a Slater determinant. For simplicity we assume a fully spin-polarized wave-function such that, we can separate off the spin component of the full physical wave-function and thus find 
\begin{eqnarray}\label{eqa.7}
	\Psi_e(\bi{r}_1,...,\bi{r}_N)=\frac{1}{\sqrt{N!}}\;\begin{tabular}{|c c c c|}
		$F_{1}(\bi{r}_1)$ &$ F_2(\bi{r}_1)$ & $\cdot\cdot\cdot$ & $F_{N}(\bi{r}_1)$\\
		$F_1(\bi{r}_2)$ & $F_2(\bi{r}_2)$ & $\cdot\cdot\cdot$ & $F_{N}(\bi{r}_2)$\\
		$\cdot$ & $\cdot$ & $\cdot\cdot\cdot$& $\cdot$\\
		$\cdot$ & $\cdot$ & $\cdot\cdot\cdot$& $\cdot$\\
		$F_1(\bi{r}_N)$ & $F_2(\bi{r}_N)$ & $\cdot\cdot\cdot$ & $F_N(\bi{r}_N)$
	\end{tabular}\quad .
	\label{tab:my_label}
\end{eqnarray}
Every component of the Slater determinant is given by a normalized mollifier
\begin{eqnarray}\label{eqa.8}
	F_i(\bi{r}_j)=
	\begin{cases}
		\mathcal{N}\exp[-\frac{1}{1-|\bi{r}_j-\bi{a}_i|^2}]\qquad \textrm{if} \qquad|\bi{r}_j-\bi{a}_i|<1 \\
		0\qquad \textrm{if} \qquad |\bi{r}_j-\bi{a}_i|\geq 1  \\
		\textrm{where}\qquad \bi{a}_i=[a+3(i-1)]\bm{\kappa}
	\end{cases}
\end{eqnarray}
where $\mathcal{N}$ is the normalization constant. The mollifiers are put on a grid along an arbitrary vector $\bm{\kappa}$ as shown in Fig.~\ref{fig:three}. The mollifiers are non-zero only within the unit ball $|\bi{r}_j-\bi{a}_i|< 1$, their supports are disjoint and the vector $\bi{a}_i$ is the center of each of these unit balls. Further, the grid of mollifiers depends on an arbitrary parameter $a$.
For the photonic part we use
\begin{equation}\label{eqa.9}
	\Phi_p(p_1,..,p_{M})=\bigotimes^{M}_{\alpha=1}\frac{1}{\sqrt{2}}(\phi_1(p_{\alpha})+\phi_2(p_{\alpha})),
\end{equation}
where $\phi_n(p_{\alpha})$ are the normalized eigen-functions of the corresponding harmonic oscillator. Thus, the full wave-function is
\begin{equation}\label{eqa.10}
	\Psi=\Psi_e(\bi{r}_1,...,\bi{r}_N)\otimes\Phi_p(p_1,..,p_{M}).
\end{equation}
Due to $\langle p_{\alpha}\Phi_p,p_{\beta}\Phi_p\rangle=2\delta_{\alpha\beta}$ we have
\begin{eqnarray}\label{eqa.11}
	\braket{\hat{V}_{int}\Psi}{\hat{V}_{int}\Psi} &=&\sum\limits^{M}_{\alpha,\beta=1}\sum\limits^{N}_{i,j=1}\braket{p_{\alpha}\Phi_p}{p_{\beta}\Phi_p} \braket{\left(\bm{\lambda}_{\alpha}\cdot\bi{r}_i\right)\Psi_e}{\left(\bm{\lambda}_{\beta}\cdot\bi{r}_j\right)\Psi_e}
        \\
        &=&2\sum\limits^M_{\alpha=1}\sum\limits^N_{i,j=1}\braket{\left(\bm{\lambda}_{\alpha}\cdot\bi{r}_i\right)\Psi_e}{\left(\bm{\lambda}_{\alpha}\cdot\bi{r}_j\right)\Psi_e}<\infty. \nonumber
\end{eqnarray}
and thus $\Psi$ can be taken as part of the domain of $\hat{H}_L$ as well as $\hat{H}'$. The expression of the energy is
\begin{eqnarray}\label{eqa.13}
\braket{\Psi}{\hat{H}^{'} \Psi}=\braket{\Psi_e}{\hat{T}_e\Psi_e}+\braket{\Psi_e}{\hat{W}_e\Psi_e}+\braket{\Psi_e}{\hat{V}_{ext}\Psi_e} +\braket{\Phi_p}{\hat{H}_p\Phi_p}+\braket{\Psi}{\hat{V}_{int}\Psi} .
\end{eqnarray}
The kinetic energy of the electrons,
\begin{eqnarray}\label{eqa.14}
\braket{\Psi_e}{\hat{T}_e\Psi_e}&=&-\frac{\hbar^2}{2m}\sum\limits^{N}_{i=1}\braket{\Psi_e}{\bi{\nabla}^2_{i}\Psi_e}\nonumber\\
&=&-\frac{\hbar^2}{2m}\sum\limits^{N}_{i=1}\prod^{N}_{n=1}\;\int \limits_{|\bi{r}_{n}-\bi{a}_i|<1}d^3r_n \Psi_e(\bi{r}_1,...,\bi{r}_N) \bi{\nabla}^2_i \Psi_e(\bi{r}_1,...,\bi{r}_N),
\end{eqnarray}
is translationally invariant and thus we can perform a change of coordinates $\bi{r}_n \longrightarrow \bi{r}_n+a\bm{\kappa}$. The integration volume after this transformation becomes
\begin{equation}\label{eqa.16}
	|\bi{r}_{n}-\bi{a}_i|<1 \longrightarrow |\bi{r}_n -3(i-1)\bm{\kappa}|<1,
\end{equation}
and thus the result of the integral is a finite constant independent of the parameter $a$ 
\begin{eqnarray}\label{eqa.17b}
 \braket{\Psi_e}{\hat{T}_e\Psi_e}=-\frac{\hbar^2}{2m}\sum\limits^{N}_{i=1}\prod^{N}_{n=1}\;\;\int \limits_{|\bi{r}_{n}-3(i-1)\bm{\kappa}|<1}d^3r_n \Psi_e(\bi{r}_1,...,\bi{r}_N) \bi{\nabla}^2_{i} \Psi_e(\bi{r}_1,...,\bi{r}_N)=A.
\end{eqnarray}
The same holds for the interaction energy, which after the same translation becomes independent of the parameter $a$
\begin{equation}\label{eqa.19}
	\braket{\Psi_e}{\hat{W}_e\Psi_e}=\sum\limits^{N}_{i<j}\prod^{N}_{n=1}\;\int \limits_{|\bi{r}_{n}-3(i-1)\bm{\kappa}|<1}d^3r_n  W_e(\bi{r}_i-\bi{r}_j)|\Psi_e(\bi{r}_1,...,\bi{r}_N)|^2=D<\infty.
\end{equation}
Without loss of generality we choose the external potential to be negative such that
\begin{equation}\label{eqa.22}
	\braket{\Psi_e}{\hat{V}_{ext}\Psi_e}=\sum\limits^N_{i=1}\braket{\Psi_e}{v_{ext}(\bi{r}_i)\Psi_e}=-\tilde{V}_a\qquad \textrm{where}\qquad \tilde{V}_a\geq 0.
\end{equation}
The energy of the photons is
\begin{eqnarray}\label{eqa.23}
\braket{\Phi_p}{\hat{H}_p\Phi_p}&=&\bigotimes^{M}_{\alpha,\beta=1} \braket{\frac{1}{\sqrt{2}}(\psi_1(p_{\alpha})+\psi_2(p_{\alpha}))}{\hat{H}_p\frac{1}{\sqrt{2}}(\psi_1(p_{\beta})+\psi_{2}(p_{\beta}))}\nonumber\\
	&=&\frac{1}{2}\sum^{M}_{\alpha=1}\left(E^{(\alpha)}_1+E^{(\alpha)}_{2}\right),
\end{eqnarray}
where $E^{(\alpha)}_n=\hbar\omega_{\alpha}(n+\frac{1}{2})$ are the eigen-energies of the harmonic oscillator. The contribution of the bilinear interaction between the electrons and the photon modes is
\begin{eqnarray}\label{eqa.24}
\braket{\Psi}{\hat{V}_{int}\Psi}&=&-\sum^{M}_{\alpha=1}\braket{\Phi_p}{p_{\alpha}\Phi_p} \braket{\Psi_e}{\left(\bm{\lambda}_{\alpha}\cdot\bi{R}\right)\Psi_e}\nonumber\\
&=&-\sum\limits^{N}_{i=1}\sum\limits^M_{\alpha=1}\bm{\lambda}_{\alpha}\braket{\Phi_p}{p_{\alpha}\Phi_p}\;\prod^{N}_{n=1}\;\;\int \limits_{|\bi{r}_{n}-\bi{a}_i|<1}d^3r_n \;\bi{r}_i|\Psi_e(\bi{r}_1,...,\bi{r}_N)|^2\nonumber\\
&=&-\sum^N_{i=1}\sum^M_{\alpha=1}\bm{\lambda}_{\alpha}\cdot\;\prod^{N}_{n=1}\;\;\int \limits_{|\bi{r}_{n}-\bi{a}_i|<1}d^3r_n \;\bi{r}_i|\Psi_e(\bi{r}_1,...,\bi{r}_N)|^2.
\end{eqnarray}
In equation (\ref{eqa.24}) we used that $\sum\limits^{M}_{\alpha=1}\bm{\lambda}_{\alpha}\braket{\Phi_p}{p_{\alpha}\Phi_p}=\sum\limits^{M}_{\alpha=1}\bm{\lambda}_{\alpha}$. We perform now again the coordinate transformation $\bi{r}_n \longrightarrow \bi{r}_n+a\bm{\kappa}$ and we have
\begin{eqnarray}\label{eqa.25}
	 \braket{\Psi}{\hat{V}_{int}\Psi}&=&-\sum^N_{i=1}\sum^M_{\alpha=1}\bm{\lambda}_{\alpha}\cdot\prod^{N}_{n=1}\;\;\int \limits_{|\bi{r}_{n}-3(i-1)\bm{\kappa}|<1}d^3r_n (\bi{r}_i+a\bm{\kappa})|\Psi_e(\bi{r}_1,...,\bi{r}_N)|^2\\
	 &=&-\sum^N_{i=1}\sum^M_{\alpha=1}\bm{\lambda}_{\alpha}\cdot\prod^{N}_{n=1}\;[\int \limits_{|\bi{r}_{n}-3(i-1)\bm{\kappa}|<1}d^3r_n \;\bi{r}_i|\Psi_e(\bi{r}_1,...,\bi{r}_N)|^2 \\
	 &+&a\bm{\kappa}\int\limits_{|\bi{r}_{n}-3(i-1)\bm{\kappa}|<1}d^3r_n |\Psi_e(\bi{r}_1,...,\bi{r}_N)|^2 ].
\end{eqnarray}
The two integrals in the above equation do not depend in the parameter $a$. The result of the first integral is some finite constant and the result of the second integral is one because it is the norm of the electronic wave-function. Consequently we obtain
\begin{eqnarray}\label{eqa.26}
	\braket{\Psi}{\hat{V}_{int}\Psi}=-B-\sum^M_{\alpha=1}\bm{\lambda}_{\alpha}\cdot\sum^{N}_{i=1}a\bm{\kappa}=-B-aN\sum^{M}_{\alpha=1}\bm{\lambda}_{\alpha}\cdot \bm{\kappa}.
\end{eqnarray}
Thus, if we choose $\bm{\kappa}$ parallel to at least one of the coupling-strength polarization vectors $\bm{\lambda}_{\alpha}$ the contribution of the bilinear interaction is proportional to $-a$, where $a$ is the position of the first mollifier in the grid we constructed.
Finally, we sum the results of all five contributions and we obtain the inequality 
\begin{eqnarray}\label{eqa.28}
	\braket{\Psi}{\hat{H}^{'}\Psi}&=& A +D-\tilde{V}_a+\frac{1}{2}\sum^{M}_{\alpha=1}\left(E^{(\alpha)}_1+E^{(\alpha)}_{2}\right)-B-aN\sum^M_{\alpha=1}\bm{\lambda}_{\alpha}\cdot\bm{\kappa
	} \leq A +D \\
	&+&\frac{1}{2}\sum^{M}_{\alpha=1}\left(E^{(\alpha)}_1+E^{(\alpha)}_{2}\right) -B -aN\sum^M_{\alpha=1}\bm{\lambda}_{\alpha}\cdot\bm{\kappa}\sim -a.
\end{eqnarray}
Since the parameter $a$ can be chosen arbitrarily we can lower the energy as much as we want. Thus, the Hamiltonian $\hat{H}^{'}$ is unbounded from below and no ground-state exists without the dipole self-energy.

\end{document}